\documentclass[aps,twocolumn,floatfix,longbibliography]{revtex4-1}
\usepackage{graphicx}
\usepackage{bm}
\usepackage{color}
\usepackage{mathrsfs}

\begin{document}
\title{Temperature--driven BCS--BEC crossover in a coupled boson--fermion system}
\author{Maciej M. Ma\'ska}
\email{maciej.maska@phys.us.edu.pl}
\affiliation{Department of Theoretical Physics, University of Silesia, Katowice, Poland}
\author{Nandini Trivedi}
\email{trivedi.15@osu.edu}
\affiliation{Department of Physics, The Ohio State University, 191 W. Woodruff Avenue, Columbus, Ohio 43210, USA}

\begin{abstract}
  We propose a simple bose-fermi model in two dimensions, with a coupling that converts pairs of opposite spin fermions into localized bosons and vice versa. We show that tracing out one of the degrees, either the bosons or fermions, generates temperature--dependent long range effective interactions between bosons as well as effective attractive interactions between fermions. Using Monte Carlo techniques we obtain the thermodynamic properties and phase stiffness as a function of temperature, dominated by vortex--antivortex unbinding of the bosons. Remarkably in the fermion sector we observe a temperature--induced BCS--BEC crossover signaled by a distinct change of their spectral properties: the minimum gap locus moves from the Fermi wave vector to the $\Gamma$ point. Such a model is relevant for describing 
  aspects of high $T_c$ superconductivity in cuprates and pnictides, superconducting islands on graphene, 
  and bose--fermi mixtures in cold atomic systems.  
\end{abstract}
\maketitle

\section {Introduction}
Strongly interacting systems lead to emergent new phases with spontaneously broken symmetries, such as magnets with broken time reversal symmetry and superconductors with broken gauge symmetry. Even within the superconducting phase the system can show effects of interactions evolving from a BCS regime with large Cooper pairs compared to inter-particle spacing to a strongly coupled regime where the Cooper pairs are tightly bound \cite{randeria-taylor}. Such an evolution of a system from a BCS to BEC regime has been predicted theoretically as a function of increasing pairing interaction and observed in experiments \cite{crossover_expts}. In this paper we propose such a BCS to BEC crossover at a fixed coupling strength but with increasing temperature. We believe this is the first such prediction originating from an explicit calculation of a model.

The original motivation for our model comes from recognizing that strongly interacting systems, such as fractional quantum Hall effect, frustrated magnets and high--$T_c$ superconductors, can be described in terms of emergent degrees of freedom that interact via fluctuating gauge fields \cite{fradkin}. In high temperature superconductors, for example, strong on--site Mott interactions generate effectively a two-component system in which the fermionic holes become superconducting in a matrix of bosonic fluctuations of the spin singlets. While many issues are still hotly debated, such as the role of intertwined order, it is nevertheless remarkable that the broad phenomenology can be understood in terms of a two-component response: (a) the hole density determining superconducting transition temperature $T_c$ and (b) the spin singlets generating the pseudogap scale $T^\ast$ below which a soft gap opens up in the density of states \cite{AtoZ}.  

Below, we investigate the spectral properties of a simple boson-fermion (BF) model as a function of coupling and temperature. The BF model, proposed in the 1950s in the context of superconductivity 
\cite{PhysRev.96.1442}, describes itinerant fermions hybridizing with bosons composed of
pairs of tightly bound opposite--spin fermions. The idea reoccurred in 1980s
in the context of electrons interacting with local lattice deformations \cite{Ran-Robasz}
and high temperature superconductivity
\cite{PhysRevLett.74.4027,PhysRevB.40.6745,PhysRevB.55.3173,PhysRevB.67.134507,PhysRevB.64.064501}. Recently, the boson--fermion
model has been adopted to describe resonance superfluids in the BCS--BEC crossover regime
\cite{Shin2008}. The model is rather general and can be applicable to a host of other realizations such as graphene with superconducting islands, and bose-fermi mixtures in cold atoms.

\medskip

\noindent{\it Model:}
The Hamiltonian of the BF model is given by
\begin{eqnarray}
H_{\rm BF}&=&-t\sum_{\langle ij\rangle,\sigma}\hat{c}^\dagger_{i\sigma}\hat{c}_{j\sigma}
+g\sum_i\left(\hat{b}^\dagger_i\hat{c}_{i\uparrow}\hat{c}_{i\downarrow}+{\rm H.c.}\right)
\nonumber \\
&-&\mu\left(2n^b+n^f\right)+E_Bn^b
\label{boson-fermion}
\end{eqnarray}
where $\hat{c}_{i\sigma}$ ($\hat{b}_i$) are fermionic (bosonic) annihilation operators,
$n^f$ ($n^b$) is the concentration of fermions (bosons), $\mu$ is the chemical potential,
and $E_B$ is the bosonic level. In the following we tune the model parameters $\mu$ 
and 
$E_B$ to have the average value $n^f=1$. We set the hopping integral $t$ as the energy unit ($t=1$).
The bosons are assumed to be localized, 
so their kinetic energy is neglected. We assume that the number of bosons per lattice site is large which allows us to approximate $\hat{b}_i\:\rightarrow\:\sqrt{n_i^{\rm b}}e^{i\theta_i}$. We then fix the number of bosons, but fermions are in the grand canonical ensemble with the chemical potential used to control their concentration.

\begin{figure}[ht]
\includegraphics[width=0.49\textwidth]{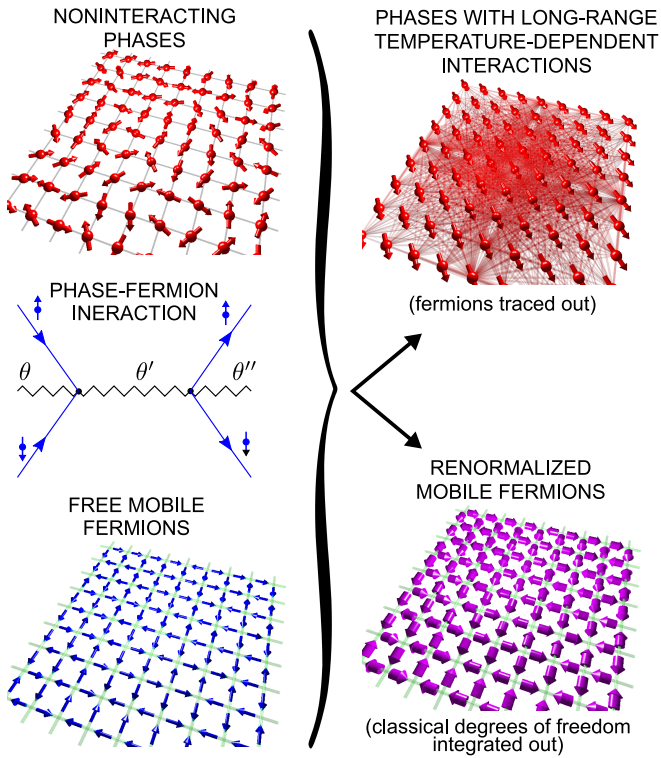}
\caption{(color online) A schematic of the phase--fermion model: mobile fermions with opposite spin scatter off bosonic phases that leads to
(a) long--range temperature--dependent interactions between the classical phases (after tracing out the fermions); and (b) renormalized itinerant fermions that show a 
change in the spectral properties as the system undergoes a BCS--BEC temperature--driven transition (after integrating out the classical degrees of freedom). 
}
\label{scheme}
\end{figure}

The BF model does not include a direct interaction between bosons. However,
through the interaction with mobile fermions, effective boson--boson
interaction gets mediated.  
Correspondingly, the properties of the fermions are also affected by their coupling to the
bosons. Since the state of the bosons is
strongly temperature dependent, the fermion--boson interaction leads to 
nontrivial behavior of the fermions, most strikingly, a temperature-induced BCS--BEC crossover.
Fig.~\ref{scheme} shows the main idea of the proposed approach.

To perform a numerical study of the model in Eq.~(\ref{boson-fermion}) we trace out the fermionic degrees of freedom,
and obtain an effective boson--boson interacting Hamiltonian. We next determine the thermodynamics of this model using Monte Carlo (MC) simulations. What distinguishes this problem from standard MC applications is that each MC step for the bosons requires diagonalizing the fermionic Hamiltonian. Such a fully quantum problem is restricted to small systems and does not allow us to access the phase transitions in the bosonic sector.  
This explains why we approximate the bosonic operators
by $c$-numbers $\hat{b}_i\:\rightarrow\:\sqrt{n_i^{\rm b}}e^{i\theta_i}$ so that quantum fluctuations of the bosonic fields are neglected while still retaining the classical phase fluctuations.

For the sake of simplicity, we assume that the boson density is uniform $n_i^{\rm b}= n^{\rm b}$,
however, this assumption is not crucial and density fluctuations can be taken into account
within the proposed approach (using, e.g., grand--canonical MC methods \cite{GCMC1,GCMC2}).
The resulting Hamiltonian describes itinerant fermions interacting with classical
phases $\theta_i$ [phase--fermion (PF) model]:
\begin{equation}
H_{\rm PF}=-t\sum_{\langle ij\rangle,\sigma}\hat{c}^\dagger_{i\sigma}\hat{c}_{j\sigma}
+\tilde{g}\sum_i\left(e^{i\theta_i}\hat{c}_{i\uparrow}\hat{c}_{i\downarrow}+{\rm H.c.}\right), 
\label{phase-fermi}
\end{equation} 
where the effective coupling includes the density of bosons $\tilde{g}\equiv g\sqrt{n^{\rm b}}$.

\medskip

\noindent{\it{Experimental Systems:}}
We propose systems 
that can be directly described by the effective Hamiltonian (\ref{phase-fermi}).

\noindent (a) The first one is graphene decorated with an array of superconducting islands
\cite{PhysRevLett.104.047001,Han2014,Eley2012}. For islands
sufficiently close to each other, Cooper pairs can tunnel directly between them and the system
can be described as 
an array of Josephson junctions. In such a system as the temperature drops 
below a Berezinsky-Kosterlitz-Thouless (BKT) transition temperature 
\cite{Berezinski1971,0022-3719-6-7-010}, the phases on the islands get 
locked and become phase coherent. The BF model suggests that even if the distance between the islands is larger
than the range of the superconducting proximity effect, phase ordering in the entire array could occur mediated by the (normal) carriers in graphene that are sensitive to the phases of the superconducting islands. 
Because of the relatively large size of the superconducting islands, the fluctuations of the
number of Cooper pairs can be neglected and the PF model (\ref{phase-fermi}) is clearly applicable.

\noindent (b) Another example is to take a 3D Bose-Einstein condensate 
of tightly bound Cooper pairs (molecules)
and break it up into a 2D array of 1D tubes, by using an optical lattice 
applied along two dimensions. 
For a sufficiently deep lattice potential, the phases between the tube-BECs are uncorrelated, hence the system as a whole is insulating. As the lattice depth is decreased, tunneling between the tubes can drive long range phase coherence across the entire system. In addition to the bosons, if unpaired fermions are present, they will experience the phase fluctuations arising from the tube-BECs. 
If the lattice potential can localize the molecules but allow the
unpaired fermions to tunnel between the BECs, the latter can mediate the 
interaction between the bosons leading to long range coherence of the isolated bosons.


\section{Bosonic Sector: \\
Effective classical Hamiltonian}
\begin{figure*}[!t]
  a)\hspace*{1mm}\includegraphics[width=0.27\textwidth]{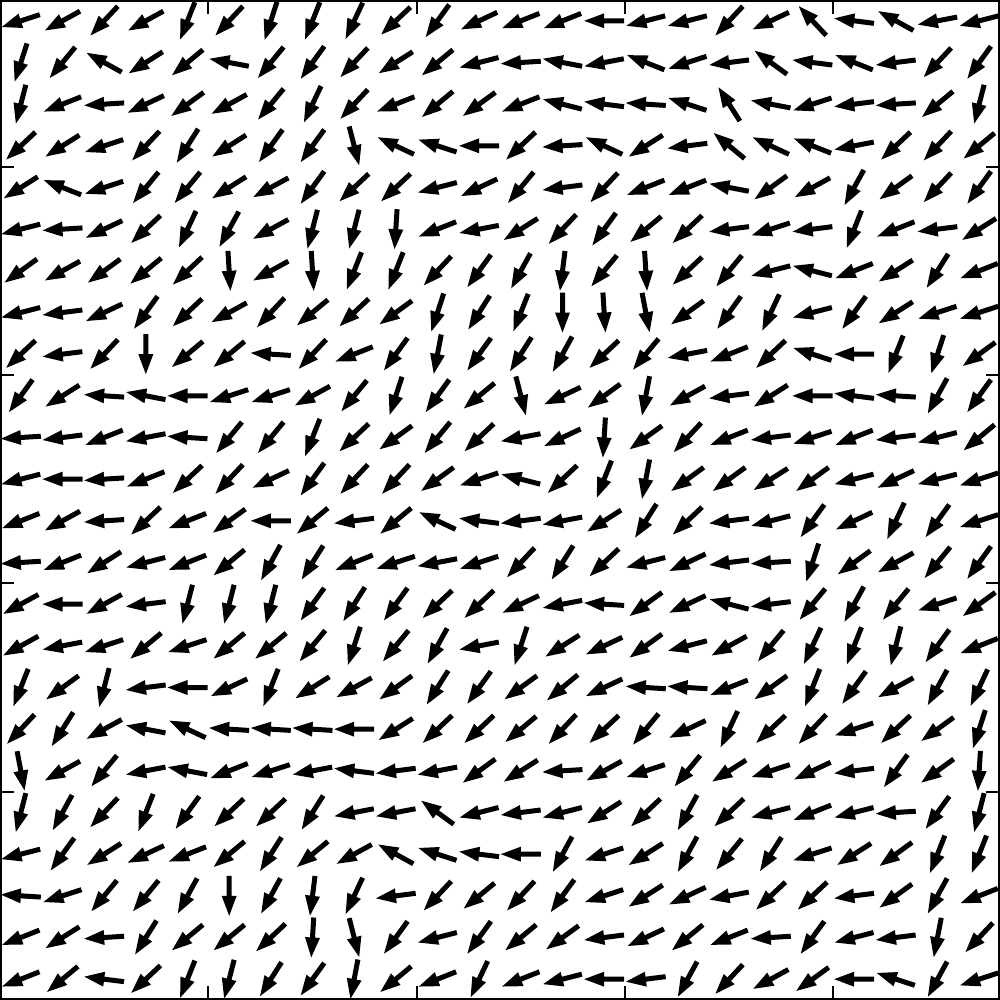}\hspace*{2mm}
  b)\hspace{1mm}\includegraphics[width=0.27\textwidth]{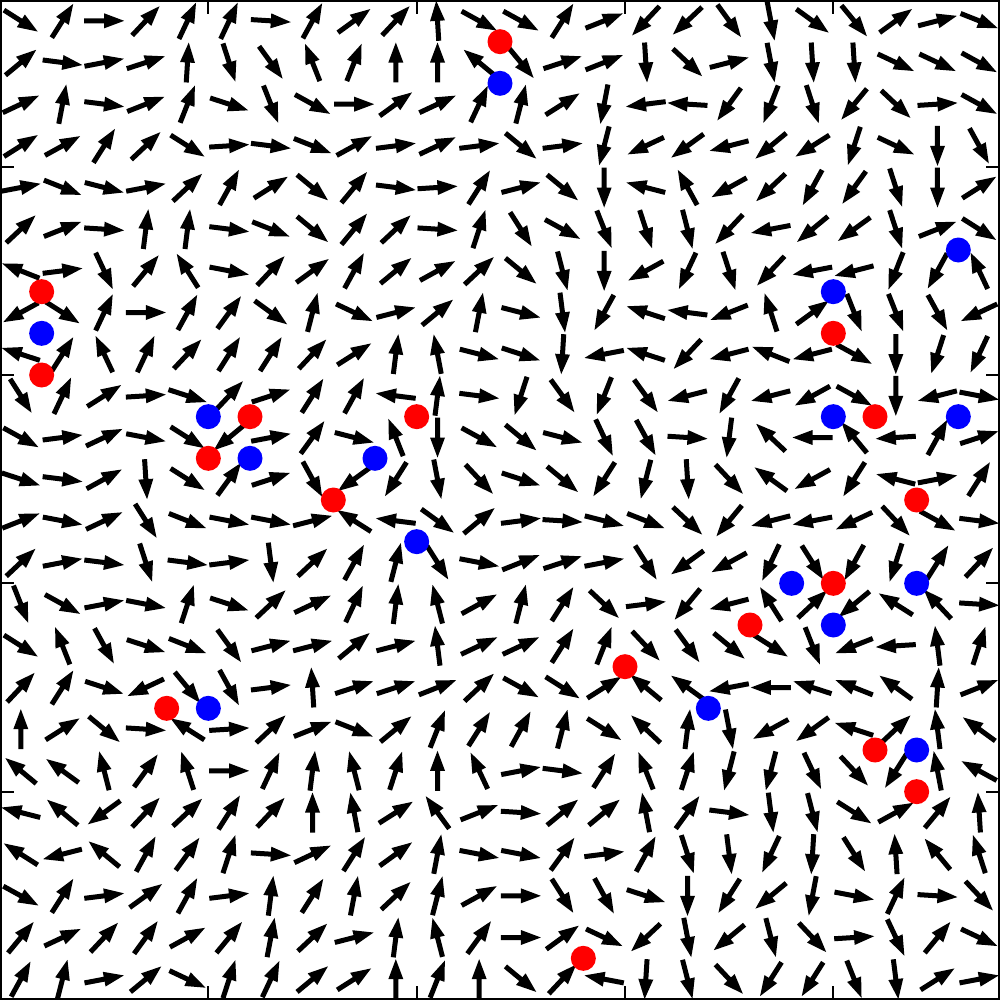}\hspace*{2mm}
  c)\hspace*{1mm}\includegraphics[width=0.27\textwidth]{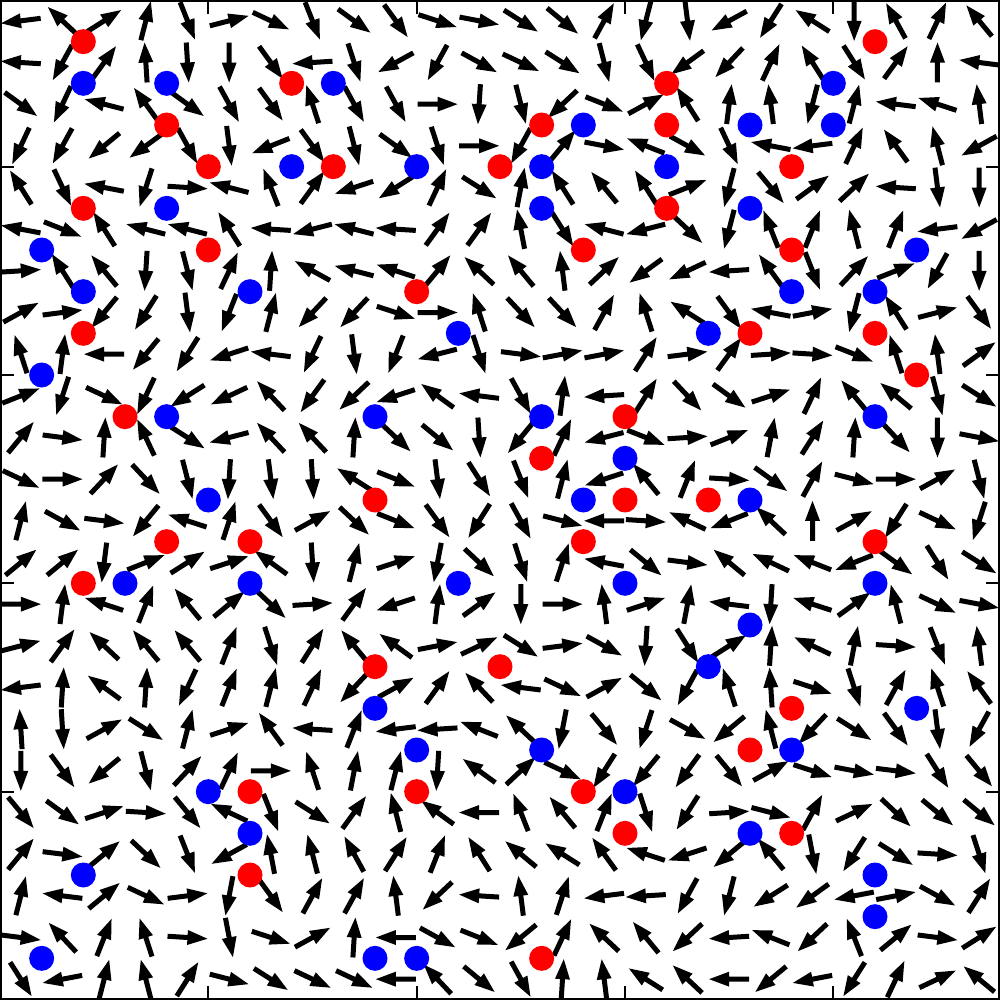}
  \caption{``Snapshots'' of phase configurations at temperatures $0.04$ (a), $0.12$ (b),
    and $0.22$ (c). Blue (red) circles represent vortices (antivortices) and the coupling $\tilde{g}=4$.}
  \label{fig2}
\end{figure*}

The Hamiltonian (\ref{phase-fermi}) represents a system described by both 
quantum (itinerant fermions) and classical (phases) degrees of freedom.
We generalize the approach in Ref.~\cite{PhysRevB.74.035109}
proposed for the Falicov-Kimball model to study the thermodynamics of the system here.
The partition function of the PF model is a sum over all possible configurations of phases and a trace over the fermionic degrees of freedom
\begin{equation}
Z=\sum_{\{\theta_i\}}{\rm Tr}\:e^{-\beta H_{\rm PF}(\vec{\theta})}=\sum_{\{\theta_i\}}e^{-\beta{\cal H}(\vec{\theta})},
\label{part-fun}
\end{equation}
where the second term is obtained after numerically tracing out the fermions exactly
for a given configuration of phases $\vec{\theta}\equiv(\theta_1,\ldots,\theta_N)$. 
Here ${\cal H}$ is a classical {\em temperature--dependent} ``Hamiltonian" 
\begin{equation}
{\cal H}(\vec{\theta})=-\frac{1}{\beta}\sum_n\ln\left[1+e^{-\beta E_n(\vec{\theta})}\right],
\label{class_ham}
\end{equation}
and $E_n(\vec{\theta})$ are its single-particle eigenvalues for
a given set of 
$\theta$'s. Note that ${\cal H}(\vec{\theta})$ is the free energy of the fermionic subsystem.
The model explicitly takes into account the time scales separating fast fermion dynamics,
treated in a fully quantum way, from the slow phase dynamics described by the classical
degrees of freedom ${\theta_i}$.

Some of the questions we address about the PF model in the bosonic sector are: (1) What are the effective interactions between bosons generated by tracing out the fermions? (2) Is there a phase transition from a low temperature ordered state of boson phases to a high temperature disordered state? (3) Is the transition described by vortex-antivortex unbinding and is it in the BKT universality class?
(4) Are there any differences in the phases or the transitions in the PF model and the standard BKT transition in the XY model with nearest neighbour interactions? In the discussion below, we answer these questions.

\begin{figure*}[ht]
    \includegraphics[width=\textwidth]{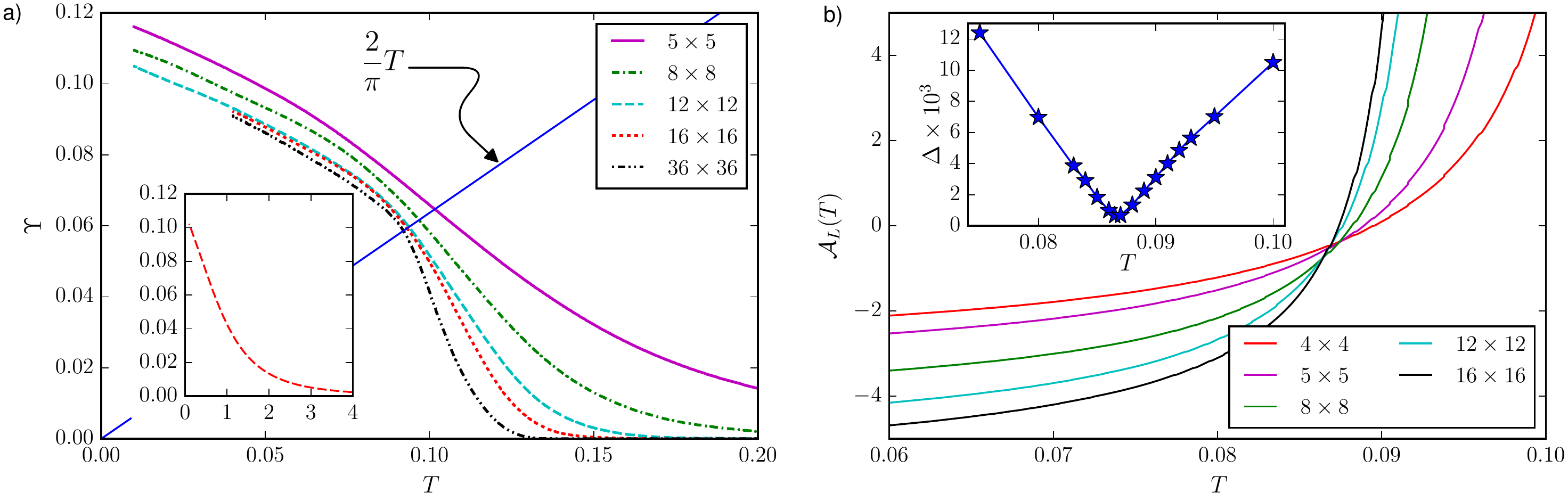}
    \caption{(color online) a) The temperature dependence of the phase stiffness $\Upsilon$ for 
    $\tilde{g}=4$ for different
    system sizes. $\Upsilon$ for the $36\times 36$ system is obtained by using the traveling cluster approximation
    TCA \cite{TCA}. The straight blue line is the BKT result $\Upsilon=\frac{2}{\pi}T$ for locating the transition. The inset shows $\Upsilon$ for a 
    $16\times 16$ system when only fermions are affected by temperature (see text). b) The temperature dependence of the l.h.s. of Eq. (\ref{c-crossings}) [${\cal A}_L(T)$] for
    different system sizes. The inset shows the root-mean-square error for fitting Eq. (\ref{RG}) to the MC results.}
  \label{stiffness}
\end{figure*}

A simple analysis of two sites (see Appendix \ref{2-site-model}) shows that in the weak coupling regime the effective interaction potential between bosons or phases deviates from the expected $\cos(\Delta\theta)$ of the usual XY model not only in the functional form but also in that the interaction $V$ is strongly temperature dependent. We present the limiting behavior for $V$ here (the full interaction potential is given in the Appendix \ref{2-site-model}), given by
\begin{equation}
V(\Delta\theta,T)\to\left\{\begin{array}{ll}
A(T)(\Delta\theta)^2& \mbox{ for }0\le \Delta\theta < \pi, \\[2ex]
A(T)(\Delta\theta-2\pi)^2& \mbox{ for }\pi\le \Delta\theta < 2\pi,
\end{array}\right.
\label{limit_hT}
\end{equation}
at low temperature and
\begin{equation}
V(\Delta\theta,T)\to A(T)\cos(\Delta\theta) 
\label{limit_lT}
\end{equation}
at high temperature.
The parameter $A(T)\to 0$ for $T/t\to\infty$ indicating that the interaction strength decreases with temperature (see Fig.~\ref{2sites_pot}). 
For stronger coupling $\tilde{g}$ the angle dependence is very well described by $\cos(\Delta\theta)$ though the coupling strength is strongly temperature dependent. 

To identify the phase transition we
follow the original approach by Fisher \cite{PhysRevA.8.1111} and calculate the response of the system to a phase twist. Specifically, we fix $\theta_i=0$ at the left edge of the system and $\theta_i=\pi$ at the right one and obtain the superfluid stiffness $\Upsilon$ from
\begin{equation}
  \Upsilon(\beta)=\frac{2}{\beta\pi^2}\int_0^\beta\left[\langle\bar{\mathscr E}(\pi)\rangle_{\beta'}
    -\langle\bar{\mathscr E}(0)\rangle_{\beta'}\right]d\beta',
  \label{stiff1}  
\end{equation}
where $\langle\ldots\rangle$ denotes an average over the phase configurations
$\vec{\theta}$ generated in MC sampling
\begin{equation}
  \langle\bar{\mathscr E}(\phi)\rangle_\beta\equiv \frac{1}{Z}\sum_{\{\theta_i\}}
  \bar{\mathscr E}\left(\vec{\theta},\phi\right)
  e^{-\beta\bar{\mathscr E}\left(\vec\theta,\phi\right)}
  \label{th_aver}
\end{equation}
and $\bar{\mathscr E}(\vec{\theta},\phi)$ is the usual quantum 
mean value of the Hamiltonian for a given
configuration $\vec{\theta}$
\begin{equation}
\bar{\mathscr E}\left(\vec\theta,\phi\right) \equiv \sum_n\left[E_n(\vec\theta,\phi)
  -\mu\right] f\left[E_n(\vec\theta,\phi)-\mu\right]. 
  \label{q_aver}
\end{equation}
$E_n(\vec\theta,\phi)$ are eigenvalues for a difference of phases on opposite 
edges equal to $\phi$ and $f(\ldots)$ is the Fermi--Dirac distribution function.

Fig.~\ref{fig2}a shows the increase of phase fluctuations with increasing temperature, the formation of vortex-antivortex bound pairs (Fig. \ref{fig2}b), and their unbinding at a
higher temperature (Fig. \ref{fig2}c).

Figure \ref{stiffness}a shows the behavior of the superfluid stiffness $\Upsilon$.
With increasing system size $\Upsilon$ gets steeper, but the
systems are still too small to display a signature of the universal jump in the stiffness.
On the other hand, in the XY model $\Upsilon$ converges rapidly at low temperature even for very small systems \cite{1742-5468-2013-09-P09001}.
The strong size dependence of $\Upsilon$ 
in the PF model 
at low temperatures can be attributed to the fact that the inter--boson coupling is long--ranged as it is generated by the itinerant fermions that travel through the entire system.

\begin{figure*}[ht]
  \includegraphics[width=\textwidth]{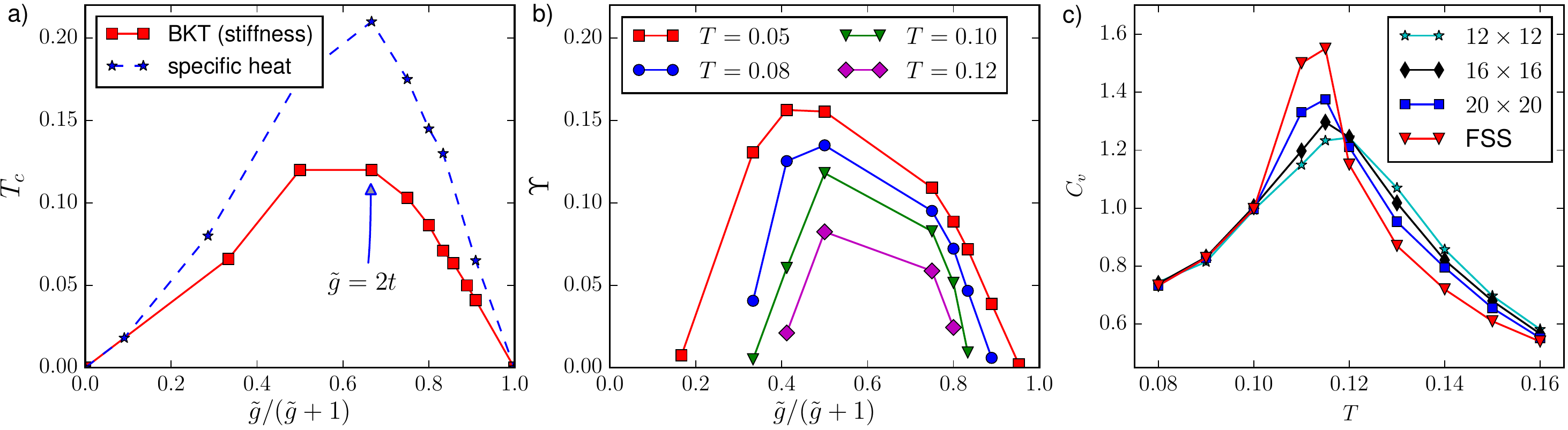}
  \caption{(color online) a) The BKT critical temperature (red solid line) and the position of the specific heat maximum (blue dashed line) as a function
    of $\tilde{g}/(\tilde{g}+1)$ in order to show the weak and strong coupling regimes compactly. The specific heat is obtained by using the dissipation--fluctuation theorem. The green line marked as $T_\Delta$ shows characteristic points in the temperature dependence of the fermionic pairing amplitude, what will be discussed in Sec.~\ref{sec:fermions}.
    b) Stiffness $\Upsilon$ as a function of the coupling $\tilde{g}$ for 
    different temperatures. c) Specific heat for $\tilde{g}=4$ as a function of 
    temperature. The red line marked FSS shows the result of the finite-size 
    scaling. In all panels the lines are only a guide to 
    the eye.}
  \label{ph-diag}
\end{figure*}
 
Given the stronger size dependence and the inability to observe a jump of the stiffness, we explore a different method to locate $T_c$. The stiffness at the BKT transition $T_c$ in the thermodynamic limit is given by \cite{0022-3719-7-6-005} $\Upsilon_{L\to\infty}=\frac{2}{\pi}T_c$.
From the knowledge of the Kosterlitz renormalization-group 
scaling at $T_c$ \cite{PhysRevB.37.5986},
\begin{equation}
  \Upsilon_{L}=\Upsilon_{L\to\infty}\left[1+\frac{1}{2}\frac{1}{\ln (L) + C}\right],
  \label{RG}
\end{equation}
where $C$ is a constant given by: 
\begin{equation}
\left\{\displaystyle\frac{2}{\pi} \int_0^{\beta}\left[\langle\bar{\cal H}(\pi)\rangle_{\beta'}
  -\langle\bar{\cal H}(0)\rangle_{\beta'}\right]d\beta'-2\right\}^{-1}-\ln(L)=C.
\label{c-crossings}
\end{equation}
Since $C$ does not depend on the system size, if one plots the l.h.s. of Eq. (\ref{c-crossings}) [denoted as ${\cal A}_L(T)$]
as a function of temperature, different sizes cross
at the same point for $T=T_c$, as shown in Fig.~\ref{stiffness}b. 
The vanishing of the fitting errors and crossings ${\cal A}_L(T)$ for all $L$
at $T=T_c$ indicates that the itinerant--fermions-mediated interaction between
the classical phases indeed drives a phase transition in the bosonic sector that is in the BKT universality class. 

The inset in Fig.~\ref{stiffness}a demonstrates that the stiffness reduction close to the BKT temperature is due to phase fluctuations and is not because of the fermions. It shows $\Upsilon$ calculated from Eq.~(\ref{stiff1}), but under the assumption that the phases are fully ordered and frozen and temperature affects the system only through the Fermi-Dirac distribution function [see Eq.~(\ref{q_aver})]. One can see that in this case the stiffness remains finite to temperatures more than an order of magnitude higher than when phase fluctuations are allowed.

By finding the critical temperature for different values of $\tilde{g}$ we construct the phase diagram shown in Fig.~\ref{ph-diag}a, where the solid red and dashed blue lines represent $T_c$ and the
position of the specific heat maximum, respectively.
For weak coupling $T_c$ increases with $\tilde{g}$, 
reaches a maximum at $\tilde{g}\approx 2$ and decreases in the strong-coupling
regime.

With increasing temperature, the rapid increase of the number of vortices
is accompanied by a maximum of the specific heat that becomes more and more pronounced
as the system size increases. The maximum, however, does not
necessarily indicate a phase transition. 
It is known that in the XY model the position
of the specific heat maximum is about 10\% above the actual critical temperature.
Fig.~\ref{ph-diag}a shows that for the PF model the separation in temperature between the specific heat maximum and the BKT transition is much larger.

\section{fermion sector\label{sec:fermions}}
It has been shown in the previous section that upon tracing out the fermionic degrees of freedom, the PF Hamiltonian leads to a model with effective interactions between phases on different sites. At sufficiently low temperatures the system develops long range phase coherence in the bosonic subsystem. As the temperature is increased the phases fluctuate 
and undergo a BKT transition at $T_c$. In this section, we analyze how the development of phase coherence in the bosonic sector affects the spectral properties of fermions. 

\medskip

\begin{figure*}[ht]
\includegraphics[width=\textwidth]{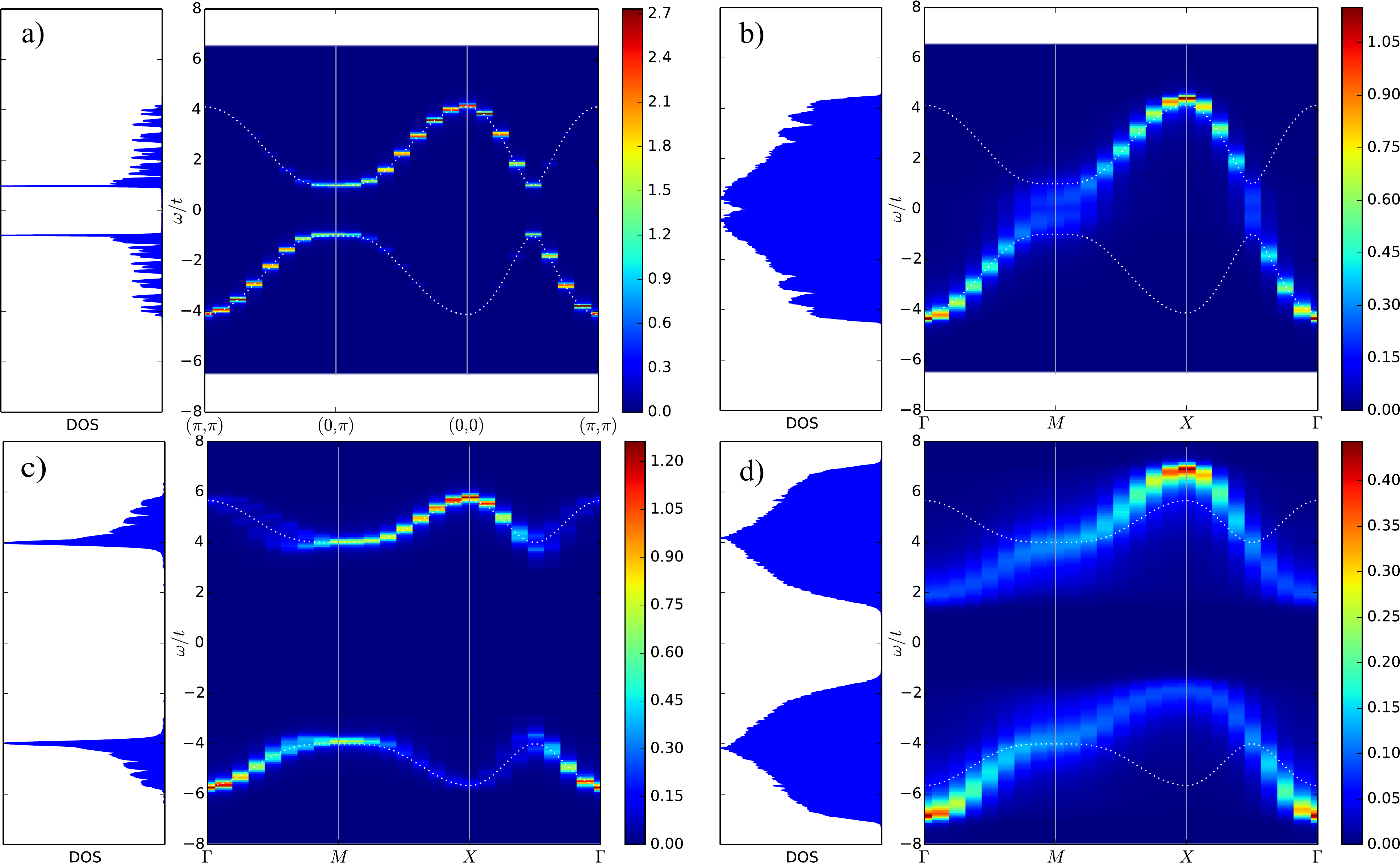}
  \caption{(color online) Densities of states and spectral functions. 
  Upper row for $\tilde{g}=1$: (a) $T=0.01$; (b) $T=1$.
  Lower row for $\tilde{g}=4$: (c) $T=0.01$; (d) $T=1$.
    The dotted lines show the dispersion of BCS quasipartlicles
    $E_{\bm k}=\pm\sqrt{\varepsilon_{\bm k}^2+\tilde{g}^2}$.}
  \label{spectr-g}
\end{figure*}

\noindent {\it Spectral function:}
For a given configuration of the bosonic phases $\vec{\theta}$ the
real-space fermionic Green function is given by
\begin{equation}
  {\cal G}(\bm{R}_i,\: \bm{R}_j,\:z)=\left\{z-{\cal H}(\vec{\theta})\right\}^{-1}_{ij},
\end{equation}
where ${\cal H}(\vec{\theta})$ is given by Eq. (\ref{class_ham}). The spectral
function is defined as
\begin{equation}
A({\bm k},\omega)=-\frac{1}{\pi}{\rm Im}\,G\left({\bm k},\omega+i0^+\right),
\end{equation}
where
\begin{equation}
  \sum_{{\bm R}_i}\sum_{{\bm R}_j} e^{i\left({\bm k}\cdot {\bm R}_i - {\bm k'}\cdot
    {\bm R}_j\right)}\langle{\cal G}(\bm{R}_i,\: \bm{R}_j,\:z)\rangle
=G({\bm k},z)\delta({\bm k}-{\bm k'}).
\end{equation}
$\langle\ldots\rangle$ denotes averaging over configurations $\vec{\theta}$ generated
in the MC sampling. By this procedure the classical bosonic degrees of freedom are integrated out
and we can analyze the fermion spectral functions and density of states (Figs. \ref{spectr-g}) renormalized by their coupling to the phases.
%
The case of $\tilde{g}=1$
($\tilde{g}=4$) represents a situation where the interaction is
weaker (stronger) than the interaction leading to maximum $T_c$ in Fig.~\ref{ph-diag}a
at intermediate coupling.

We find that for weak coupling, $\tilde{g}=1$,
the system behaves like a BCS superconductor as expected with a gap that closes
at $T_c$. However, for strong coupling, $\tilde{g}=4$, the spectral function at low temperatures is BCS--like with a minimum gap at the Fermi energy.
However upon increasing the temperature there are two interesting observations: (a)
the gap does not close even for arbitrarily high
temperatures; and (b) the spectral function looks BEC--like with a minimum gap moving to the $\Gamma$ point.
A similar change in the spectral function has been reported in the attractive Hubbard model on two coupled triangular lattices \cite{PhysRevX.6.021029} at $T=0$. 
We report here the first evidence for a temperature--driven phase transition from a BCS superconductor to a disordered bosonic insulator. 
The mechanism responsible for this transition can be illustrated with the PF model on only two sites, and provides useful insight; see Appendix \ref{2-site-model}.

Essentially, temperature affects the fermionic spectrum primarily via the bosonic phases. At low temperatures, fermions interact with ordered phases and the spectral lines are relatively narrow. At high temperatures, the fermions are scattered by the disordered phases that changes their momenta and energies and leads to broadening of the spectral lines. Note here that the broadening due to the Fermi--Dirac distribution function is negligible and does not affect their shape.

\medskip

\begin{figure*}[ht]
  \includegraphics[width=\textwidth]{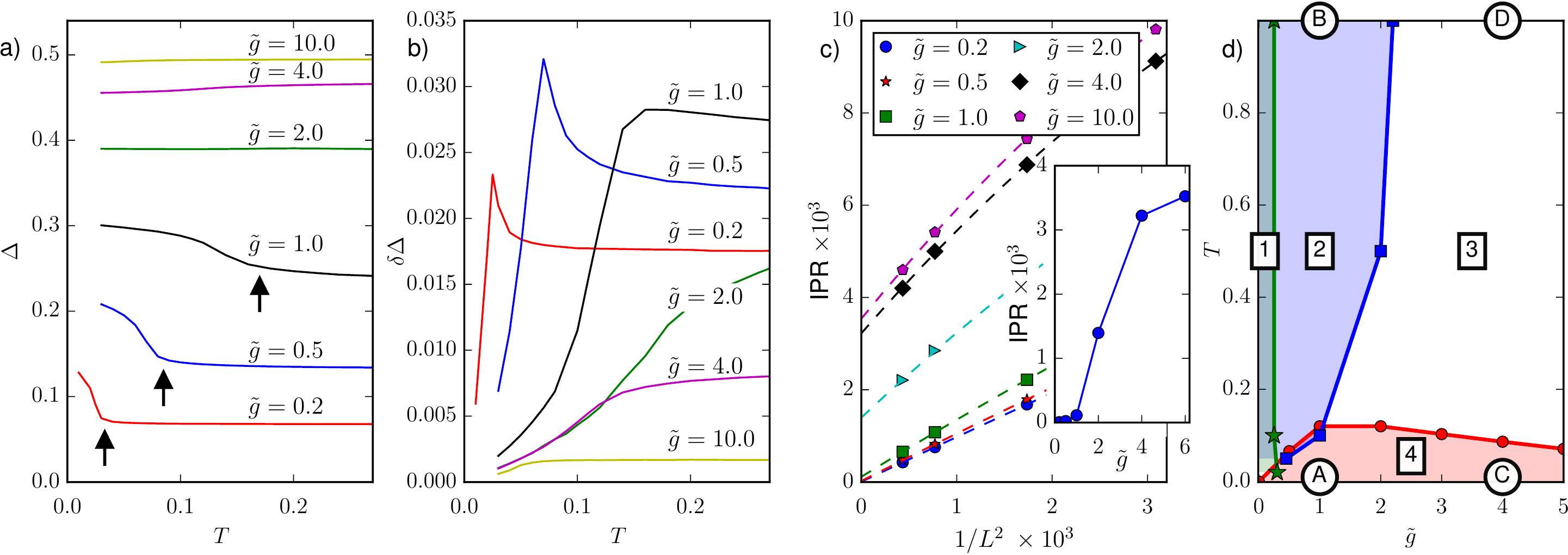}
  \caption{(color online) a) Temperature dependence of $\Delta$ given by Eq.~(\ref{eq:delta}) for different coupling constants. The characteristic points 
  marked with arrows are denoted in the text as $T_\Delta$.
  b) Fluctuations of $\Delta$.
  c) The finite size scaling of the IPR at $T=1$. The inset shows the value of the IPR extrapolated to the thermodynamic limit. d) The phase diagram of the phase--fermion model. The spectral functions corresponding
    to letters A, B, C and D
    are presented in Figs. \ref{spectr-g}a, \ref{spectr-g}b,
    \ref{spectr-g}c, and \ref{spectr-g}d, respectively. The regions marked with numbers
    are: 1 -- fermionic delocalized states, 2 -- fermionic states with a gapless spectrum,
    3 -- bosonic insulator, 4 -- phase coherent bosons (finite stiffness). 
  }
  \label{delta}
\end{figure*}

\noindent{\it Pairing amplitude:}
The difference between the weak and strong coupling regimes is seen also in the magnitude of induced pairing between the fermions. Fig.~\ref{delta}a shows the dependence of the pairing amplitude 
\begin{equation}
\Delta\equiv\frac{1}{N}\sum_i\langle c_{i\uparrow}c_{i\downarrow}\rangle
\label{eq:delta}
\end{equation}
on temperature.
In weak coupling $\Delta$ decreases with increasing temperature, whereas in strong coupling regime it remains almost temperature independent, slightly increasing for $\tilde{g}$ around $4$. In both cases at some characteristic temperature $T_\Delta$ (indicated in Fig.~\ref{delta}a), the pairing amplitude $\Delta$ reaches a constant value and does not change with further increase of temperature.  For strong coupling $T_\Delta$ is difficult to determine because of its weak temperature dependence. In weak coupling $T_\Delta$ tracks the specific heat maximum, as shown in Fig.~\ref{ph-diag}a. Additionally, in this limit fluctuations of $\Delta$ 
are strongly pronounced at $T_\Delta$, as shown in Fig.~\ref{ph-diag}b.

We attribute the difference in the temperature dependence of $\Delta$ to the balance of different contributions to the total energy: in weak coupling the fermionic pairs are weakly bound and with the increase of temperature, it is energetically favorable to break up the pairs and gain the kinetic energy for the unpaired fermions. As a result, $\Delta$ decreases with increasing temperature.

On the other hand, for strongly bound pairs it is preferable to stay in the paired state even at the cost of localization of the pair wave functions. This makes the pairing amplitude almost temperature independent. The system nevertheless undergoes a temperature--driven superconductor--insulator transition due to the vanishing of the superfluid stiffness.

Analogous to the spectral functions, the temperature dependence of $\Delta$ is a result of the phase disorder and not merely the consequences of broadening of the Fermi-Dirac distribution function.

\medskip

\noindent {\it{Mobility of fermions:}}
The properties of fermions are connected to their mobility. In order to find the regime where the single fermion wave functions are localized we calculate the {\em inverse participation ratio} (IPR), a quantity that in the thermodynamic limit tends to zero for delocalized states and remains finite for localized ones, defined by:
\begin{equation}
  {\rm IPR}=\frac{\sum_{ni}|\Psi_{ni}|^4}{\sum_{ni}|\Psi_{ni}|^2},
  \label{eq:ipr}
\end{equation}
where $\Psi_{ni}$ is amplitude of the $n$-th state at site $i$. In order to perform finite size scaling we simulate large systems up to $48\times 48$ using the
Travelling Cluster Approximation \cite{TCA,PhysRevE.91.063303}. 

Our results are presented in Fig. \ref{delta}c. Above the BKT transition,
fermions are delocalized only for very weak coupling. The transition between localized and delocalized states is almost independent of temperature. Our calculations of the density of states at the Fermi level
shows that the fermions get localized while the energy gap is still closed. We
attribute the localization in this regime to interactions with disordered phases. There exists
a unitary transformation of the fermionic operators that transfers the bosonic phases to
the hopping integral $t\to t\exp\left[i(\theta_i-\theta_j)/2\right]$ which shows that the
localization of fermions can be understood also as the localization due to a random magnetic field
\cite{PhysRevLett.82.604}.

With increasing coupling a gap opens in the spectrum. For $\tilde{g}\le 1$ the gap opens at the
BKT temperature; for larger $\tilde{g}$, however, the temperature
at which the gap opens increases, whereas the $T_{\rm BKT}$ remains constant up to $\tilde{g}=2$
and then decreases. The overall phase diagram is presented in Fig. \ref{delta}d.

\section{Summary}
We have proposed an effective boson--fermion model where the interaction between the classical and quantum degrees of freedom leads to highly nontrivial properties. In particular, we have shown that fermions mediate an effective interaction between bosonic phases that is long--ranged and temperature--dependent. This interaction is significantly different from that in the 2D XY model, but nevertheless also leads to a phase transition in the BKT universality class. The bosonic phase configurations fluctuate with temperature and affect the properties of fermions.
At low temperatures the phases are almost aligned uniformly and their interaction
with fermions drives a BCS--type pairing in the fermionic subsystem. However, when temperature increases, the bosnic phases become disordered and leads to localization of the fermions. When the coupling of the fermions with the bosonic phases is strong, 
the fermions are still paired, but in this regime they form a BEC state with different spectral properties: the minimum gap locus shifts from the Fermi surface to the $\Gamma$ point. we thus find a temperature--driven BCS--BEC transition. 

We have so far assumed in the PF Hamiltonian that the bosons are localized. By allowing them to hop between lattice sites one can introduce a direct interaction that competes or cooperates with the interaction mediated by the fermions. 

\begin{acknowledgments}
  M.M.M. acknowledges support by National Science Centre (Poland) under Grant No. DEC-2013/11/B/ST3/0082 and N.T. acknowledges DOE-Basic Energy Sciences grant DE-FG02-07ER46423. 
\end{acknowledgments}

\appendix

\section{Two--site model\label{2-site-model}}
The PF model can be analytically solved for a system composed of two sites. 
\begin{figure}[h]
  \includegraphics[width=0.25\textwidth]{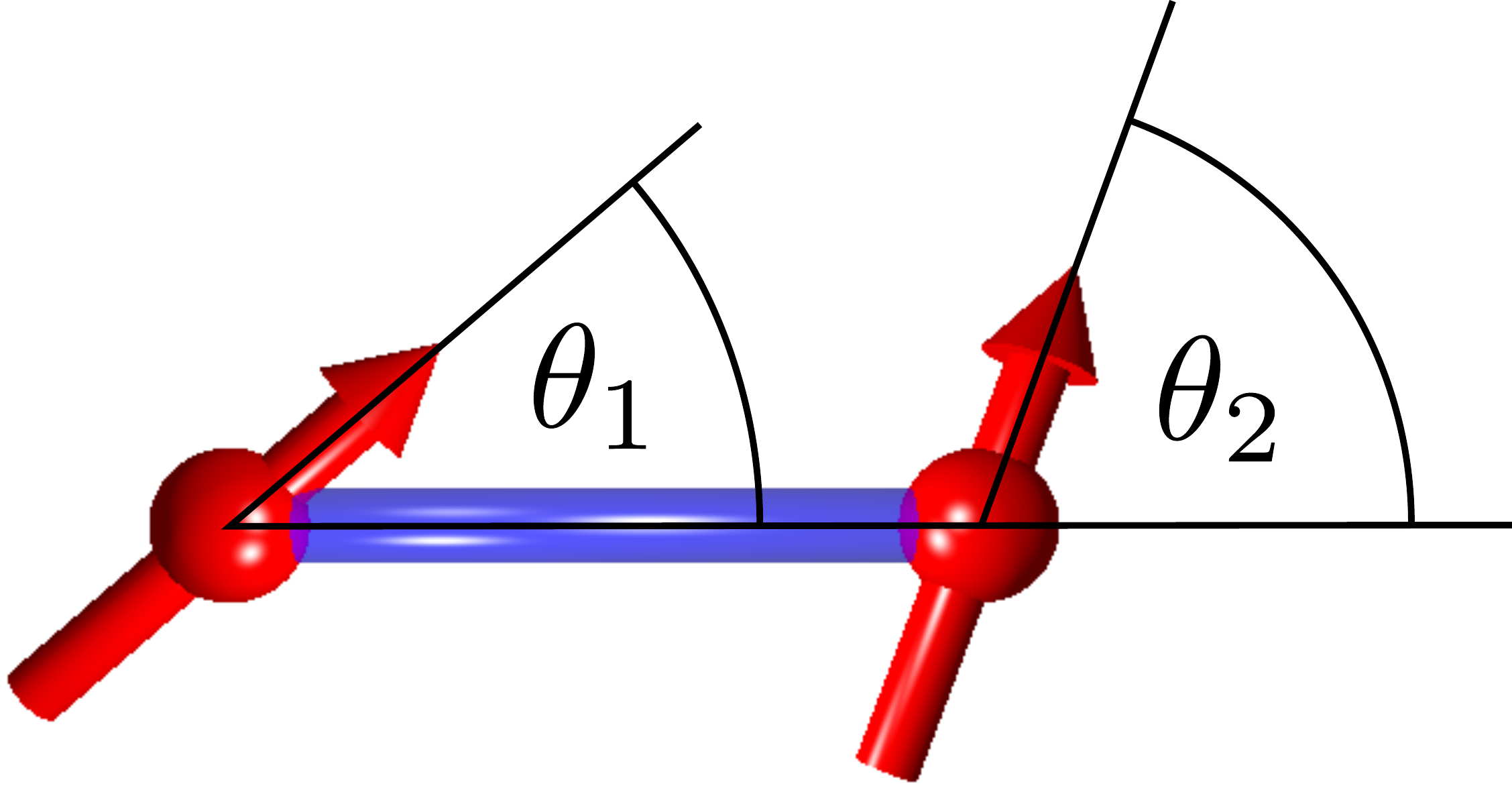}
  \caption{(color online) Two--site PF model. The thick blue line represents fermions coupled to phases $\theta_1$ and $\theta_2$.\label{2sites_fig}}
\end{figure}
The Hamiltonian is given by
\begin{eqnarray}
H(\theta_1,\theta_2)=&-&t\sum_\sigma\left(c^\dagger_{1\sigma} c_{2\sigma}+c^\dagger_{2\sigma} c_{1\sigma}\right)
\nonumber \\
&+&\tilde{g}\left(e^{i\theta_1}c_{1\uparrow}c_{1\downarrow}+e^{i\theta_2}c_{2\uparrow}c_{2\downarrow} + \mbox{H.c.}\right),
\label{2sites}
\end{eqnarray}
with eigenenergies 
\begin{eqnarray}
E_1&=&-\sqrt{t^2+\tilde{g}^2+2\tilde{g}t\left|\sin\left(\Delta\theta/2\right)\right|},\label{Eis1}\\
E_2&=&-\sqrt{t^2+\tilde{g}^2-2\tilde{g}t\left|\sin\left(\Delta\theta/2\right)\right|},\label{Eis2}\\
E_3&=&-E_1,\label{Eis3}\\
E_4&=&-E_2,
\label{Eis4}
\end{eqnarray}
where $\Delta\theta\equiv\theta_1-\theta_2$. The partition function 
\begin{equation}
  Z=\int d\theta_2\int d\theta_2 {\rm Tr}e^{-\beta H(\theta_1,\theta_2)}=2\pi\int d\Delta\theta e^{\beta V(\Delta\theta,\beta)},
\end{equation}
where the effective temperature--dependent interaction between phases $\theta_1$ and $\theta_2$ is given by
\begin{equation}
V(\Delta\theta,\beta)=-\frac{1}{\beta}\sum_{i=1}^4\ln\left[1+e^{-\beta E_i(\Delta\theta)}\right]
\label{eq:2site_pot}
\end{equation}
is equal to the free energy of the fermionic subsystem. In the weak coupling regime the potential has very different dependence on $\Delta\theta$ at low and high temperature (Fig.~\ref{2sites_pot}a). The limits of the angle-dependent part of $V$ are given by Eqs.~(\ref{limit_hT}) and (\ref{limit_lT}).
For stronger coupling the difference is much smaller and for $\tilde{g}=4$ (Fig.~\ref{2sites_pot}b) the angle dependence is very well described by $\cos(\Delta\theta)$ at arbitrary temperatures. 
Note, that in many cases $V(\Delta\theta)$ has the same form as in the XY model.
\begin{figure}[h]
  \includegraphics[width=0.48\textwidth]{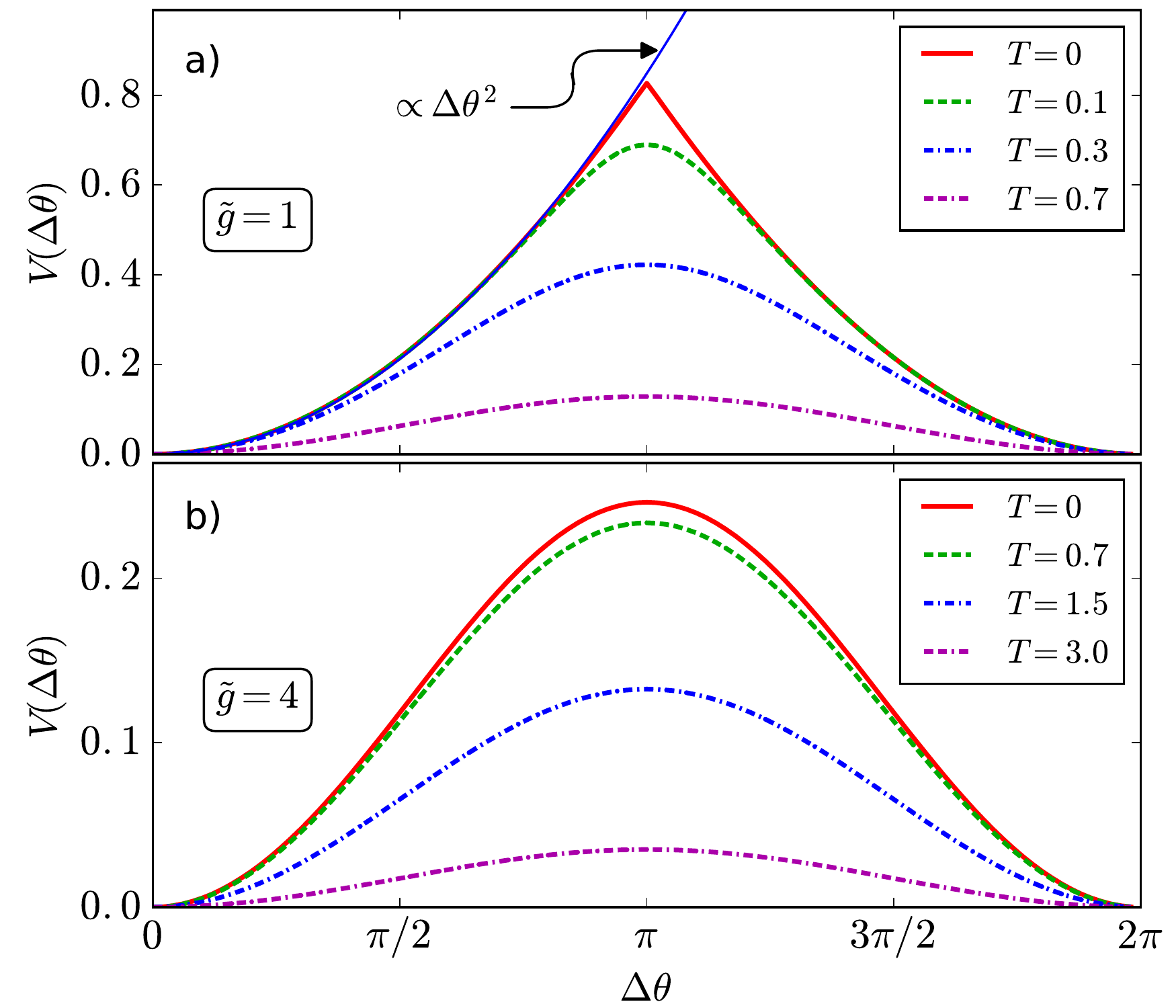}
  \caption{(color online) Phase--phase interaction potential for a two--site PF model given by Eq.~(\ref{eq:2site_pot}) for $\tilde{g}=1$ (a) and $\tilde{g}=4$ (b). The potentials have been shifted so that $V(0)=0$. \label{2sites_pot}}
\end{figure}
It can be seen in Fig.~\ref{2sites_pot} that the magnitude of the interaction strongly decreases when temperature increases.

Since in the 2D PF model fermions travel through the entire system, the phase--phase interaction that is mediated by them has long--range character. The knowledge of the interaction in a two-site model allows one to determine the importance of the long--range part of this effective interaction. To this end we have performed MC simulations for a classical model in which phases on a 2D square lattice were interacting only with their nearest neighbors with the interaction potential given by Eq.~(\ref{eq:2site_pot}). In this way the microscopic mechanism that leads to the effective phase--phase interaction is the same as in the original PF model, but the range of the interaction is limited to nearest neighbors only. Since for a wide range of the model parameters the interaction in the two--site model is well described by $\cos(\Delta\theta)$, one can expect that the BKT transition should be observed also in the classical model with only nearest--neighbor interaction. MC simulations confirms this assumptions, but it turns out that the critical temperature is {\em higher} than in the original PF model with long--range interaction. The convergence with increasing system size is similar to that presented in Fig.~\ref{stiffness} and the difference in the critical temperature is, for example, around 20\% for $\tilde{g}=4$.

\subsection{BCS--BEC Crossover}
While there is no collective behavior in the two--site model, it still can be useful to get an insight into the mechanism of the temperature--driven BCS--BEC crossover. 

\begin{figure*}[ht]
  \includegraphics[width=0.49\textwidth]{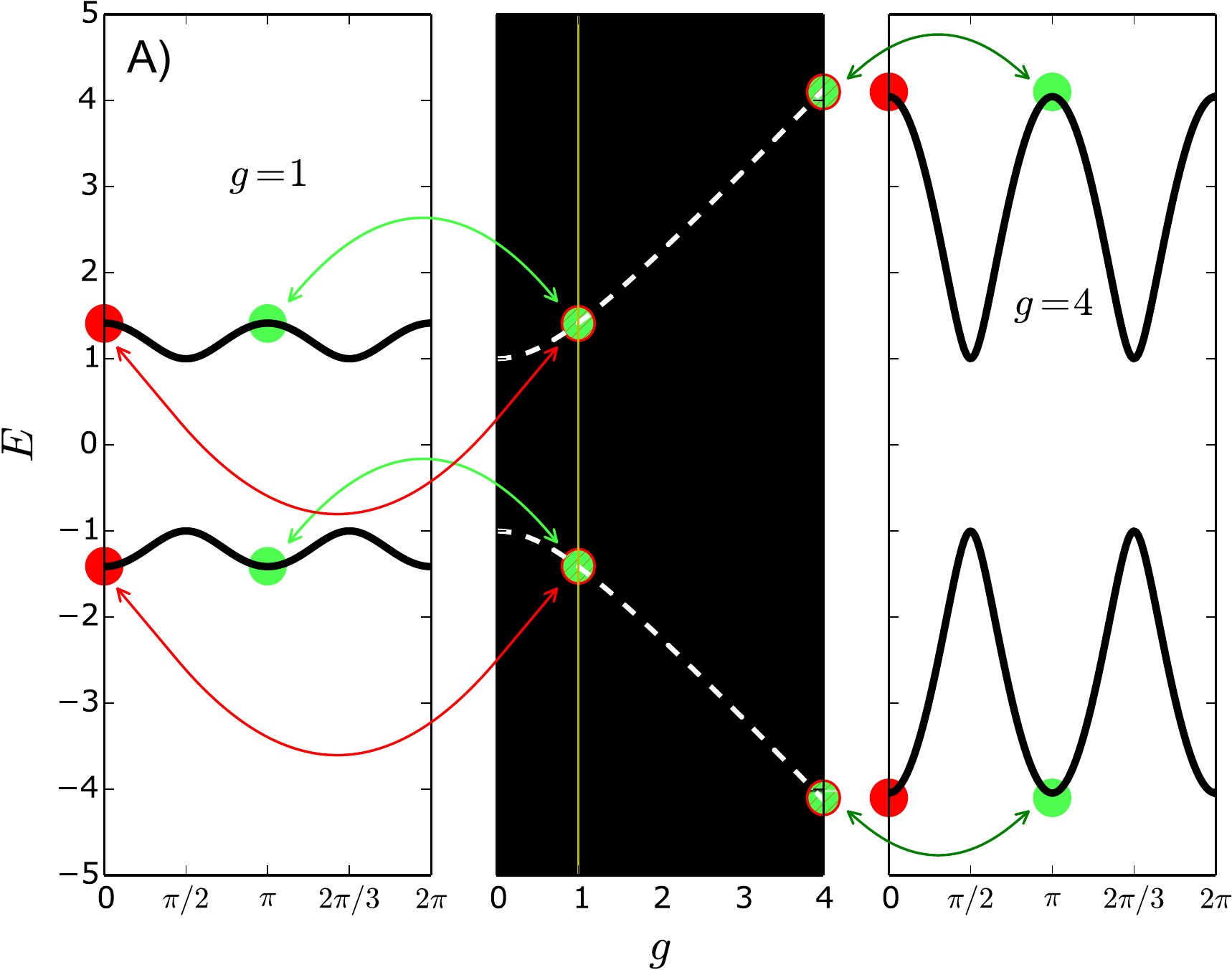}\hspace*{4mm}
  \includegraphics[width=0.49\textwidth]{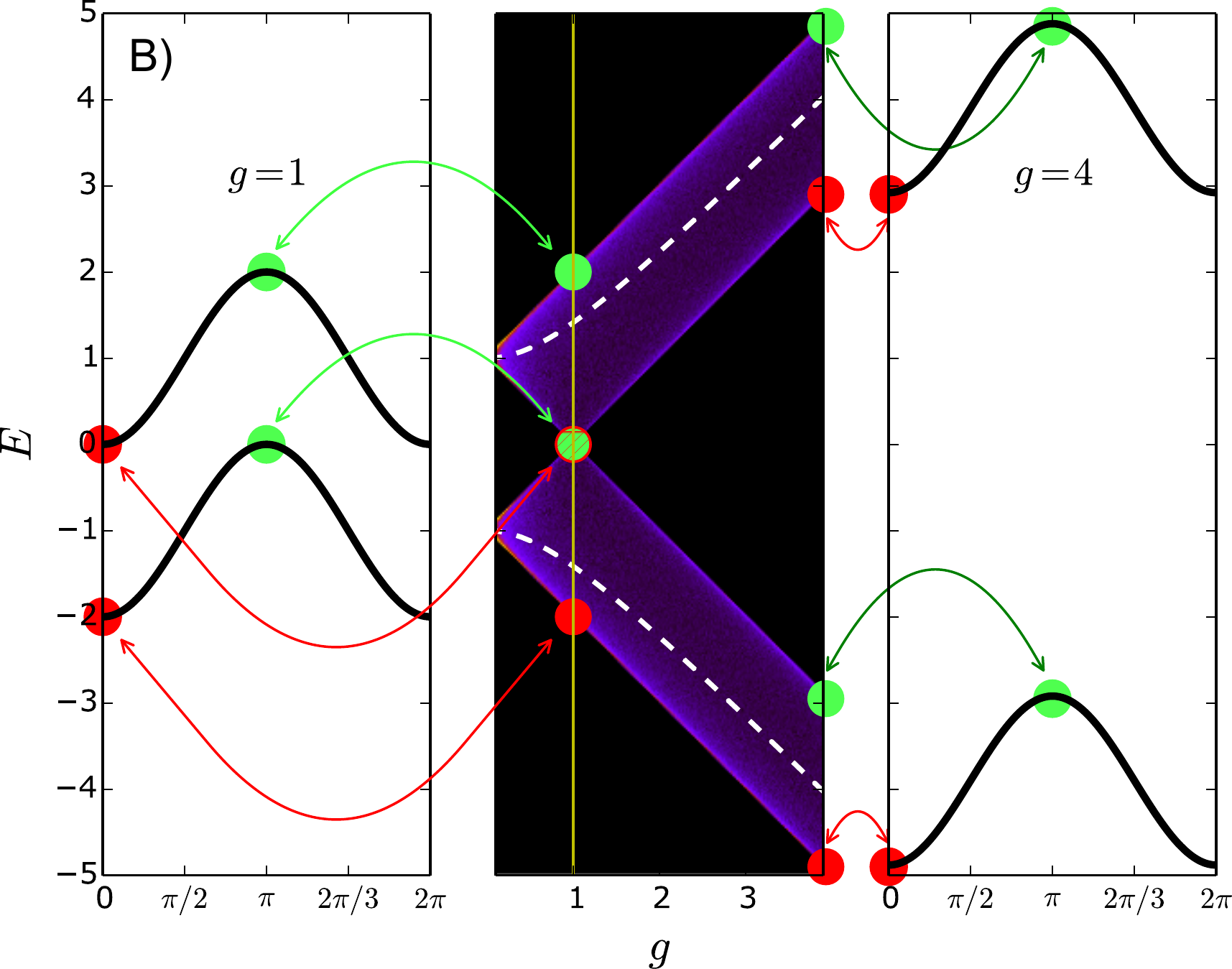}
  \caption{(color online) Spectral functions of a 2-site model for $\Delta\theta=0$ (low temperature, A) and
    for $\Delta\theta$ with uniform distribution between 0 and $2\pi$ (high temperature, B).
    In both cases the white dashed line in the middle panels show $E_i$ given by Eqs. (\ref{Eis1}-\ref{Eis4}).
    The shaded area around this line in (B) represents the corresponding distribution of
    $E_i$'s for $\Delta\theta\in [0,\:2\pi)$. The left and right panels show the dispersion
    relations $A(k)=\pm\sqrt{\varepsilon_{k}^2+\tilde{g}^2}$ (A) and
    $A(k)=\pm|\varepsilon_{k}\pm\tilde{g}|$ (B) for $\tilde{g}=1$ (left panels) and $\tilde{g}=4$ (right panels). The red (green) circles correspond to points $k=0$ ($k=\pi$). Values for which $A(0)=A({\pi})$ are represented by striped
    red-green circles. Red and green arrows show the correspondence 
between the spectral functions and the eigenvalues of the Hamiltonian.
    \label{2site_spectr}}
\end{figure*}

 At low temperature
$\Delta\theta$ is small and there are two doubly-degenerate BCS--like eigenvalues
$\pm\sqrt{\varepsilon_{k}^2+\tilde{g}^2}$ (the Brillouin zone includes only points 0 and $\pi$ and 
$\varepsilon_0=t$ and $\varepsilon_\pi=-t$). However, at high temperature $\Delta\theta$ has a
uniform distribution between 0 and $2\pi$. It can be easily shown that then the distribution
of the eigenvalues is strongly peaked at $\pm|\varepsilon_{k}\pm\tilde{g}|$. The transition
is illustrated in Fig. \ref{2site_spectr}.

\section{Vortex-antivortex interaction\label{voravor}}
\begin{figure}[h]
  \includegraphics[width=0.49\textwidth]{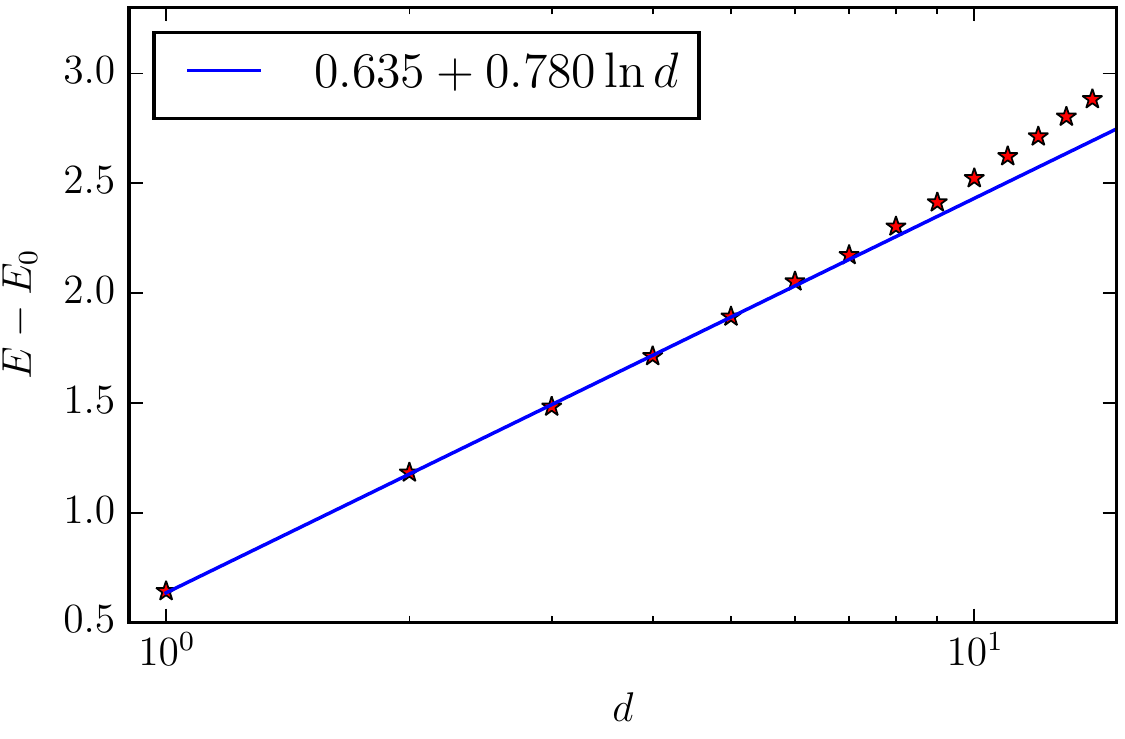}
  \caption{(color online) Vortex-antivortex interaction energy as a function of the distance between them for a $20\times 20$ system at $T=0.02$. The blue line shows a logarithmic dependence best fitted to MC results at small distances. The deviation occurs when the inter--vortex distance is comparable to the linear size of the system and the phases are affected by the finite size effects.}
  \label{vor-avor}
\end{figure}

The BKT transition results from a logarithmic dependence of the vortex-antivortex interaction on the distance between them. The same behavior can be demonstrated for the PH model. Namely, we fix phases at corners of an elementary plaquette in such way that they form a vortex. Another fixed phases form an antivortex at some distance $d$. Then we start MC simulations for all other phases and after the system gets thermalized we measure its energy $E$ with respect to the energy of a similar system without any fixed phases $E_0$. The simulations were performed at low temperature applying the simulated annealing method, so the final configuration is close to the ground state. Fig. \ref{vor-avor} shows $E-E_0$ as a function of $d$. 

For small $d$ the dependence is perfectly described by the logarithmic function. For larger $d$, when is becomes comparable to the linear size of the system, the phases are affected by the presence of the system boundaries (we used periodic boundary conditions) and the energy deviates. 

\section{Methods}
Eqs.~(\ref{part-fun}) and (\ref{class_ham}) allow us to perform the classical 
MC simulation for the PF model: in each MC step a phase in a randomly chosen 
lattice site is changed, the Hamiltonian matrix is diagonalized and the new 
configuration is accepted or rejected according to the Metropolis criterion. The difference 
between MC simulations for, e.g., the Ising model and the present one is that here we use the 
free energy of the fermionic subsystem (\ref{class_ham}) instead of the internal energy in the 
acceptance criterion. It is also much more time-consuming, because $\theta$'s are 
continuous variables, not just $\pm 1$ like in the Ising model, and for a 2D $L\times L$
system in each MC step 
$2L\times 2L$ matrix has to be diagonalized. On the other hand, the size of the Hilbert
space is still much smaller than in the case of systems with full quantum correlations.
Additionally, it is possible to significantly speed up the calculations using, e.g.,
Chebyshev expansion of the fermion density of states \cite{doi:10.1143/JPSJ.68.3853}, an algorithm that
directly updates the spectrum of a successive Hamiltonian matrix based on the spectrum of
previous Hamiltonian matrix \cite{1742-5468-2007-08-P08007}, or Restricted Boltzmann Machines \cite{RBM}. Some of the present calculations
were performed with the help of the Traveling Cluster Approximation (TCA) \cite{TCA,PhysRevE.91.063303}.

\bibliographystyle{apsrev4-1}

%

\end{document}